\documentclass{aa}
\usepackage{epsf}

\begin{document}


\newcommand{\pF}{\mbox{$p_{\mbox{\raisebox{-0.3ex}{\scriptsize F}}}$}}  
\newcommand{\vph}[1]{\mbox{$\vphantom{#1}$}}  
\newcommand{\kB}{\mbox{$k_{\rm B}$}}           
\newcommand{\vF}{\mbox{$v_{\mbox{\raisebox{-0.3ex}{\scriptsize F}}}$}}  
\renewcommand{\arraystretch}{1.5}
\newcommand{\dd}{{\rm d}}


\title{Adiabatic Index of Dense Matter
and Damping of Neutron Star Pulsations}
\author{
       P.~Haensel\inst{1}
\and
       K.~P.~Levenfish\inst{2}
\and
       D.~G.~Yakovlev\inst{2}
        }
\institute{
       N.~Copernicus Astronomical Center,
       Bartycka 18, 00-716 Warszawa, Poland\\
       {\it e-mail:} {\tt haensel@camk.edu.pl}
\and
        Ioffe Physical Technical Institute, Politekhnicheskaya 26,
        194021 St.-Petersburg, Russia \\
        {\it e-mails:} {\tt  ksen@astro.ioffe.rssi.ru, 
                        yak@astro.ioffe.rssi.ru}
}

\date{}
\offprints{P.Haensel: {\tt haensel@camk.edu.pl}}

\titlerunning{Adiabatic index and damping of neutron star pulsations}
\authorrunning{P.~Haensel et al.}

\abstract{The adiabatic index $\Gamma_1$
for perturbations
of dense matter is studied
under various physical conditions which can prevail in neutron
star cores. The dependence of $\Gamma_1$
on the composition of matter (in particular,
on the presence of hyperons), on the stellar pulsation amplitude,
and on the baryon superfluidity is analyzed. 
Timescales of damping of stellar pulsations
are estimated 
at different compositions, temperatures, and pulsation
amplitudes. 
Damping of pulsations by bulk viscosity in the neutron-star
cores can prevent the stars to pulsate
with relative amplitudes $ \ga (1-15)\%$
(depending on the composition of matter). 
}

\maketitle


\section{Introduction}

The adiabatic index $\Gamma_1=
(n_{\rm b}/P)({\rm d}P/{\rm d}n_{\rm b})$ for 
density perturbations determines the
changes of pressure $P$ associated with variations of the local baryon density
$n_{\rm b}$ (e.g., Shapiro \& Teukolsky 1983).
It enters the equations governing small-amplitude neutron-star
pulsations (Thorne \& Campolattaro 1967, Thorne 1968, 1969) as well as
the criteria of stability of cold relativistic stars (Meltzer \& Thorne 1966,
Chanmugan \& Gabriel 1971, Chanmugan 1977, Gourgoulhon et al. 1995).
For the time-dependent perturbations, like neutron-star pulsations, $\Gamma_1$
has to be calculated taking into account the slowness of various 
equilibration channels in dense matter. 
The same factors strongly affect viscous damping of
neutron-star pulsations.

In the present paper, we study  
three main factors which regulate $\Gamma_1$
and the viscous damping of pulsations:
composition of dense matter, pulsation amplitude, and superfluidity of baryons.
Equilibration processes for various compositions of neutron-star cores are
studied in Sect. 2. 
In Sect. 3, we calculate the adiabatic index under various
conditions of  density, composition, and temperature.
We separately study two
different regimes:
first, the regime with the perturbations of
chemical potentials of particles
much smaller than $T$,
and second, the regime with the perturbations
much larger than $T$ (we use the units in which
the Boltzmann constant $k_{\rm B}=1$); 
we also consider the effect of baryon
superfluidity. In Sect.\ 4 we
estimate the
timescales of the viscous damping of density pulsations under various
physical conditions which can be realized 
in the neutron-star cores. Finally, in
Sect.\ 5 we summarize our results and briefly discuss
the problems which remain to be
solved.

\section{Relaxation processes in neutron-star cores}
\label{sect:Processes}

It is widely believed that
dense matter of the outer neutron-star core
($0.5 \, n_0 \la n_{\rm b} \la 2 \, n_0$,
where $n_0 \approx 0.16$ fm$^{-3}$ is the number density
of saturated nuclear matter) contains neutrons (n),
protons (p), electrons (e), and -- if the electron Fermi energy exceeds
the muon rest energy
(that occurs at $n_{\rm b}\ga n_0$)  --  also  muons ($\mu$).
At higher densities in the inner core, hyperons
may appear, first of all, $\Sigma^-$ and $\Lambda$ hyperons.
All constituents of matter are strongly degenerate.
We restrict ourselves to this composition and
will not consider exotic models of matter containing pion or
kaon condensates or free quarks.
We will study a neutrino-transparent neutron-star 
core formed in about 30~s after
the neutron star birth in a supernova explosion.
We ignore thus the early stage of the neutrino opaque core
in a protoneutron star (the adiabatic index in such matter
was studied by  Gondek et al.\ \cite{ghz97}).

\subsection{Strong and weak-interaction processes}
\label{sect:List.processes}

Full thermodynamic equilibrium in dense matter
is established only after partial equilibria are achieved
in all equilibration channels. In our case, there may be five types
of channels [labelled as (a), (a$'$), (b), (c), and (d)],
with drastically different relaxation
times. These channels,
described partly, e.g., by Yakovlev et al.\
(\cite{ykgh01}) and Haensel et al.\ (\cite{hly02}), are
as follows:\\[0.1ex]
(a) Strong and Coulombic elastic collisions,
\begin{equation}
   {\rm BB^\prime \rightleftharpoons BB^\prime,\quad
   \ell B \rightleftharpoons \ell B, \quad \ell \ell'
   \rightleftharpoons \ell \ell',}
\label{rapid1}
\end{equation}
where B and ${\rm B}^\prime$ are baryons (nucleons B=N or
hyperons B=H), while
$\ell$ and $\ell'$ are leptons in dense matter (electrons or muons).
These processes do not change particle fractions.\\[0.1ex]
(${\rm a}^\prime$) Strong-interaction processes conserving
strangeness but changing baryon fractions,
\begin{equation}
   {\rm n \Lambda \rightleftharpoons p \Sigma^- }.
\label{rapid2}
\end{equation}
(b) Modified Urca processes like
\begin{equation}
    {\rm n N} \to {\rm N^\prime p} \ell \bar{\nu}_l ,
    \quad {\rm pN}\ell \to {\rm n N^\prime} \nu_{\ell},
\label{Murca}
\end{equation}
where $\nu_{\ell}$ and $\bar{\nu}_{\ell}$
are neutrino and antineutrino.
Similar processes can also proceed with one or both nucleons
replaced by hyperons. These processes can change the fractions
of nucleons, hyperons, and leptons.\\[0.1ex]
(c) Direct Urca processes involving nucleons and hyperons,
\begin{eqnarray}
  && {\rm n \to p}\ell\bar{\nu}_\ell, 
     \quad {\rm p}\ell \to {\rm n} \nu_\ell;
\label{Durca}\\
  && {\rm \Sigma^- \to n}\ell \bar{\nu}_\ell ,
  \quad {\rm n}\ell \to \Sigma^- \nu_\ell ;
\label{DurcaSigma}\\
  && \Lambda \to {\rm p}\ell  \bar{\nu}_\ell ,
  \quad {\rm p}\ell \to \Lambda \nu_\ell ,
\label{SurcaLambda}
\end{eqnarray}
which, similarly to processes (b),
 can change the fractions of nucleons, hyperons,
and leptons. \\[0.1ex]
(d) Nonleptonic collisions of baryons changing
strangeness, e.g.,
\begin{equation}
 {\rm   N} \Lambda  \rightleftharpoons  {\rm N n}, \quad
   {\rm nn \rightleftharpoons  p} \Sigma^- ;
\label{strange}
\end{equation}
they can change the fractions of baryons, but do not affect
the fractions of leptons.

\subsection{Relaxation times}
\label{sect:Relax.times}

Relaxation times in channels (a) and (${\rm a}^\prime$) are about
$(10^{-16}-10^{-19})/T_9^2$ s or shorter, where $T_9\equiv T/10^9~{\rm K}$
(this can be deduced, e.g., from the results of 
Flowers \& Itoh \cite{fi79}). 
Therefore, these channels guarantee instantaneous (on hydrodynamical
time scales) partial thermodynamic 
equilibration to the Fermi-Dirac distributions of
particles of all species $j$ with a temperature 
$T$ and chemical potentials $\mu_j$.
Processes (${\rm a}^\prime$) allow for additional equilibration of
$\Lambda$, $\Sigma^-$, and nucleons, which leads to
the equality of the chemical potentials
$\mu_\Sigma+\mu_p=\mu_n+\mu_\Lambda$.
The relaxation in channels (b), (c), and (d) (with relaxation
times $\tau_{\rm M}$, $\tau_{\rm D}$, and $\tau_{\rm H}$) 
is associated with weak
interactions and lasts, therefore, much longer;
it will be important for our analysis.
In a nonsuperfluid matter close to full equilibrium
the relaxation times can be estimated as (e.g., Yakovlev et al.\
\cite{ykgh01}, Haensel et al.\ \cite{hly02})
\begin{equation}
  \tau_{\rm M} \sim {2\,  {\rm month} \over T_9^6}, \quad
  \tau_{\rm D} \sim {20\, {\rm s} \over T_9^4},  \quad
  \tau_{\rm H} \sim {1 \, {\rm ms} \over T_9^2}.
%
%
\label{relaxtimes}
\end{equation}
These estimates are valid if only the pulsation amplitudes
$ \delta \mu_j \equiv
|\mu_j - \mu^{(0)}_j|$ are very low, $\delta \mu_j \ll T$,
where $\mu_j^{(0)}$ is the fully equilibrium value of
$\mu_j$. We will call this regime as the {\it regime
of subthermal pulsations}.
 
According to Eqs.\ (\ref{relaxtimes}),
the slowest are modified Urca processes which involve
neutrinos and largest number of strongly degenerate particles. 
The most rapid are nonleptonic processes
changing strangeness; they go via weak interaction
but involve smallest number of degenerate particles
and no neutrinos.
While estimating $\tau_{\rm H}$
and viscous dissipation parameters in hyperonic matter
from the results of Haensel et al.\ (\cite{hly02}), we set the
phenomenological hyperon reaction
parameter $\chi = 0.1$. All the parameters in hyperonic matter
are rather uncertain and can differ from the presented
values within a factor of ten or even larger.

\section{Adiabatic index}
\label{sect:Gamma.calc}

Clearly, one can divide the neutron star core into
three zones with different relaxation properties.
The outer zone (zone ``M'') extends from the core-crust interface,
$n_{\rm b} \sim 0.5 \, n_0$, to some baryon density
$n_{\rm D}$. The relaxation in this zone is provided by
the modified Urca processes (b). The intermediate zone
``D'' extends from the density $n_{\rm D}$,
at which the direct Urca process becomes open,
to some density $n_{\rm H}$.  The relaxation
goes much faster there,
because it is provided by the direct Urca processes (c)
while the modified Urca processes are always insignificant
once the direct Urca processes are open. The inner zone ``H''
extends from the density $n_{\rm H}$,
at which hyperonic processes
(d) start to operate,
to the stellar center.
In this zone, we have the relaxation channels of two very different types:
direct Urca (c) and nonlepton hyperonic processes (d),
with $\tau_{\rm H} \ll \tau_{\rm D}$.
The low-mass neutron stars contain only zone M.
More massive stars may contain zones D and H.

In a pulsating star, the baryon number density
$n_{\rm b}$ varies with pulsation period
$2 \pi /\omega \sim 10^{-3}$ s,
where $\omega \sim 10^4$ s$^{-1}$ is a typical
pulsation frequency. Even if the unperturbed (non-pulsating)
stellar state was in the full equilibrium
(which we assume), the pulsating matter
may be out of the equilibrium with respect to channels
(b)--(d).
The adiabatic index for perturbations of
the  matter has, therefore,  to be defined as
\begin{equation}
   \Gamma_1 = {n_{\rm b}\over P}\,
             \left( {\dd P\over \dd n_{\rm b}} \right)_{\rm C}
            = {\rho+P/c^2 \over P}\,
            \left( {\dd P\over \dd \rho} \right  )_{\rm C},
\label{DefGamma}
\end{equation}
where $\rho$ is the mass density
and the subscript C means that the derivative should be
constrained by the conditions in various channels (see below).
In the full equilibrium, we have $\Gamma_1= \Gamma_{\rm EQ}$,
where $\Gamma_{\rm EQ}$
is the adiabatic index
that determines the stiffness of the equation of state
(EOS) in a non-pulsating star (i.e., determines the
basic neutron-star structure in equilibrium).
Figure 1 shows the density dependence of
$\Gamma_{\rm EQ}$ for a relativistic mean-field EOS
of Glendenning (\cite{glendenning85}; his case 3) 
of npe$\mu \Lambda \Sigma^-$ matter (similar curve
of $\Gamma_{\rm EQ}$ was calculated, for instance,
by Balberg \& Gal \cite{bg97}).
For this EOS, $\mu$, $\Lambda$, and $\Sigma^-$ particles appear at
$n_{\rm b}=0.110$, 0.310, and 0.319 fm$^{-3}$, respectively,
and the main direct Urca process, Eq.\ (\ref{Durca}), 
with electrons is allowed at
$n_{\rm D} = 0.227$ fm$^{-3}$.
The  adiabatic index $\Gamma_1$ in question
may differ from $\Gamma_{\rm EQ}$ because of
the lack of equilibrium in channels (b)--(d) in pulsating matter.

Exact calculation of $\Gamma_1$ is complicated.
Thus we restrict ourselves to
several important limiting cases.

Generally,
the pressure depends not only on $n_{\rm b}$ but also
on the fractions $x_j=n_i/n_{\rm b}$ of various
particle species $j$.
Thus, Eq.\ (\ref{DefGamma}) can be written as
\begin{equation}
  \Gamma_1 = \Gamma_{\rm FR} + {n_{\rm b} \over P} \sum_{j} \strut
          \left({\partial  P \over \partial x_j }\right)_{n_{\rm b}}
          \left({\partial x_j \over \partial n_{\rm b} }\right)_{\rm C},
\label{Gamma}
\end{equation}
where $\Gamma_{\rm FR}=
(\partial \ln P / \partial \ln n_{\rm b})_{\{x_j\}}$
is the adiabatic index calculated for the ``frozen''
composition of matter (at $x_j=$ constant).
The frozen-composition index $\Gamma_{\rm FR}$ is also depicted in
Fig.\ 1 for the same model of dense matter.
Freezing the particle fractions makes an EOS
stiffer and increases $\Gamma_1$.
The real value of $\Gamma_1$ lies evidently
between $\Gamma_{\rm EQ}$ and  $\Gamma_{\rm FR}$.
Some examples are shown in Fig.\ 2.

\begin{figure}
\centering
\epsfxsize=8.7 cm
\epsffile[30 300 486 695]{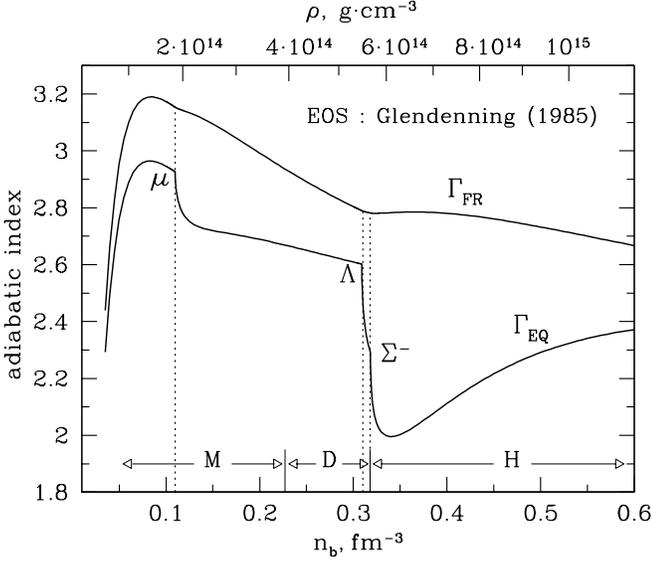}
\caption{
Adiabatic index $\Gamma_1$
versus baryon number density (lower horizontal scale) or
mass density (upper horizontal scale) 
in zones M, D, and H (indicated by arrows).
The curves show $\Gamma_{\rm EQ}$
at full equilibrium and
$\Gamma_{\rm FR}$ at fully frozen composition of matter.
Vertical dotted lines:
the thresholds of appearance of $\mu$, $\Lambda$,
and $\Sigma^-$.
}
\label{fig1}
\end{figure}

Two obvious conditions
of varying the parameters of matter
while calculating the derivatives in Eq.\ (\ref{Gamma})
are: conservation of baryon
number and electric neutrality. This reduces the number
of independent fractions (degrees of freedom) by two.
Other conditions are described below.

\subsection{Subthermal pulsations in nonsuperfluid matter}
\label{sect:small.amp.normal}

Let us consider nonsuperfluid matter and the subthermal pulsation
regime, $\delta \mu_j \la T$, for all $j$. In zones M and D,
the relaxation times $\tau_{\rm M}$ and $\tau_{\rm D}$ are much larger
than the pulsation periods $2 \pi / \omega$. Accordingly,
$\Gamma_1=\Gamma_{\rm FR}$
is a good approximation, i.e., $\Gamma_1$ is
noticeably higher than $\Gamma_{\rm EQ}$ (Figs.~1 and 2).

In zone H, the situation is more complicated.
If the star is sufficiently cold, $T \ll 10^9$ K,
we have $\tau_{\rm H} \gg 2 \pi / \omega$,
and again $\Gamma_1=\Gamma_{\rm FR}$.
However, in a hotter star, we may have the opposite
condition, $\tau_{\rm H} \ll 2 \pi / \omega$, which will lead
to the equilibrium in hyperonic
channels (d); then $\Gamma_1$ has to be calculated
assuming this equilibrium. The latter implies two
conditions:
\begin{equation}
  \mu_{\rm n} + \mu_\Lambda \! =\! \mu_{\rm p} + \mu_\Sigma,\qquad
  x_\Sigma + x_\Lambda 
       =  x_\Sigma^{(0)} + x_\Lambda^{(0)},
\label{d.cond.mu.x}
\end{equation}
in pulsating matter (the upperscript (0) refers to equilibrium values).
In combination with conservation of baryon number
and electric neutrality in pulsating matter,
conditions (\ref{d.cond.mu.x})
allow us to calculate the partial derivatives
in Eq.\ (\ref{Gamma}) and determine $\Gamma_1$
(plotted by the thick solid line in Fig.\ 2).
This adiabatic index is only slightly lower
than $\Gamma_{\rm FR}$, for the adopted EOS
(cf.\ the thick solid line and
the upper dashed line in Fig.\ 2).

\begin{figure}
\centering
\epsfxsize=8.7 cm
\epsffile[30 300 486 695]{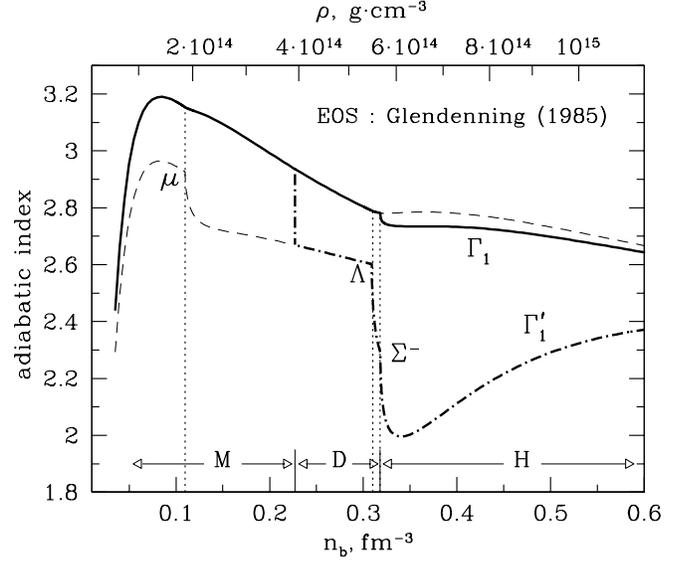}
\caption{
Same as in Fig.\ \ref{fig1} but for different equilibration regimes.
Thick solid line: $\Gamma_1$ for subthermal perturbations with
Urca channels (b) and (c) frozen, but
hyperonic channels (d) fully relaxed.
Thick dash-and-dotted line in zones D and H:
$\Gamma_1=\Gamma_1^{\, \prime}$ for
suprathermal perturbations
with all the channels, but
modified Urca, fully relaxed.
Thin dashed curves: $\Gamma_{\rm EQ}$ and
$\Gamma_{\rm FR}$ as in Fig.\ \ref{fig1}.
}
\label{fig2}
\end{figure}

\subsection{Suprathermal pulsations in nonsuperfluid matter}
\label{supra}

Let us outline the case of larger pulsation amplitudes,
$\delta \mu_j \gg T$, which however are assumed to be relatively
small, $\delta \mu_j  \ll \mu_j$, so that the linear
pulsation theory is valid.
Under such conditions, the deviations of the chemical
potentials from the fully equilibrium values will
strongly reduce the relaxation times $\tau$.
As follows from calculations of the rates of
nonequilibrium processes (e.g., Reisenegger \cite{reisenegger95},
Yakovlev et al.\ \cite{ykgh01}),
one can
estimate the relaxation time at $\delta \mu_j \gg T$
by replacing
\begin{equation}
 T \to \widetilde{T} = \sqrt{ T^2 + (\delta \mu)^2/(2 \pi)^2}
\label{newT}
\end{equation}
in Eqs.\ (\ref{relaxtimes}), where
$\delta \mu$ is a typical pulsation amplitude
of chemical potentials. This replacement correctly
reproduces the regime of subthermal pulsations
($\delta \mu \la T$, $\widetilde{T} \approx T$).
In the suprathermal regime, $\delta \mu \gg T$,
the relaxation times no longer depend on $T$
but depend on the pulsation amplitude $\delta \mu$ instead.
Let us stress that the replacement is approximate;
detailed calculations are required to obtain
quantitatively accurate results in the suprathermal
regime. Such calculations are beyond the scope of
the present paper.

To illustrate the suprathermal effects
let us take, 
for instance, $T= 100$ keV $ \sim 10^9$ K, $\mu \sim 500$ MeV,
and $\delta \mu \sim 0.02 \, \mu \sim 10$ MeV. 
Then we have
$\delta \mu/T \sim 100$ and the relaxation times
$\tau_{\rm M} \sim $ 0.1 s,
$\tau_{\rm D} \sim 10^{-4}$ s,
and $\tau_{\rm H} \sim 3 \times 10^{-6}$ s.
Thus, $\tau_{\rm M}$ remains much larger than $2\pi/\omega$,
and we have $\Gamma_1 = \Gamma_{\rm FR}$, as before, in zone M.
However, $\tau_{\rm D,H} \ll 2\pi/\omega$,
and we have rapid relaxation
in all channels in zones D and H,
where we can put $\Gamma_1=\Gamma_{\rm EQ}$.
The adiabatic index for this regime
is plotted by the dot-and-dashed line
in Fig.~2 (marked as $\Gamma^\prime_1$).

\subsection{Subthermal pulsations in superfluid matter}
\label{sect:small.sup}

Now we return to subthermal
pulsations, $\delta \mu \la T$, but assume superfluidity
of baryons. All baryon species (n, p, $\Lambda$, and
$\Sigma^-$, in our example) can be in a superfluid state.
The critical temperatures $T_{\rm c}$ of
baryon superfluids depend on the model
of strong interactions and the many-body theory employed.
Various microscopic models predict density
dependent $T_{\rm c}$, different for different
baryons, ranging from $\sim 10^8$ K to $\sim 10^{10}$ K
(e.g., Lombardo \& Schulze \cite{ls01}; also see
Haensel et al.\ \cite{hly01,hly02} for references).
A strong superfluidity ($T_{\rm c} \gg T$) will drastically
(exponentially) reduce the rates of the processes involving baryons
(due to the appearance of superfluid gaps in baryon
energy spectra; see, e.g., Yakovlev et al.\ \cite{ykgh01}).
This will
increase the relaxation times (\ref{relaxtimes}) in various
channels (b)--(d).
In this case we can easily have
$\tau \gg 2\pi/\omega$, and the relaxation will be frozen over
pulsation periods. Thus, putting $\Gamma_1=\Gamma_{\rm FR}$
will be a good approximation.

\subsection{Suprathermal pulsations in superfluid matter}
\label{supra-sup}

This case is complicated.  
Note however that the pulsations with amplitudes
$\delta \mu$ higher than the superfluid gaps, $\Delta \la 1$ MeV,
seem to be independent of superfluid properties of dense matter.
Thus, the relaxation times of
pulsations with such amplitudes
can be estimated using the results of Sect.\  
\ref{supra}.

\section{Damping of
stellar pulsations by bulk viscosity}
\label{sect:bulk.vis}

Let us turn to the bulk viscosity of dense matter
and associated damping of neutron-star pulsations. 
For the composition of dense matter
we are interested in, these problems have been
studied in a number of papers starting from the
pioneering papers by Langer \& Cameron (\cite{lc69})
and Jones (\cite{jones71}).
The current state of the problem and references to
other papers can be found in
Jones (\cite{jones01}), Haensel et al.\ (\cite{hly02}), and
Lindblom \& Owen (\cite{lo02}).

\subsection{Nonsuperfluid matter}
\label{sect:normal.zeta.nonlin}

Simple estimates of the bulk viscosity $\zeta$
in the familiar subthermal regime are given, for instance, by
Haensel et al.\ (\cite{hly01,hly02}). The generalization
to the case of suprathermal pulsations is straightforward.
Generally, one can introduce an {\it effective bulk viscosity}
$\widetilde{\zeta}$ which determines
actual dissipation rate of pulsation energy per unit volume.
The estimates of $\widetilde{\zeta}$ 
can be obtained from the estimates of $\zeta$
by replacing there $T \to \widetilde{T}$.
In this way we come to the
order-of-magnitude estimates of the effective bulk viscosity
in zones M, D, and H:
\begin{eqnarray}
    \widetilde{\zeta}_{\rm M} & \sim & 5 \times 10^{18}\, \omega_4^{-2}
                 \, \widetilde{T}_9^6 \; \; {\rm g~cm^{-1}~s^{-1}},
\label{zetaM} \\
   \widetilde{\zeta}_{\rm D} & \sim & 5 \times 10^{24}\, \omega_4^{-2}
                 \, \widetilde{T}_9^4 \; \; {\rm g~cm^{-1}~s^{-1}},
\label{zetaD} \\
    \widetilde{\zeta}_{\rm H} & \sim & 10^{29}\, \omega_4^{-2}
                 \, \widetilde{T}_9^2 \; \; {\rm g~cm^{-1}~s^{-1}},
\label{zetaH}
\end{eqnarray}
where $\omega_4$ is the pulsation frequency in units of
$10^4$ s$^{-1}$.

At the next step, following Haensel et al.\ (2002),
we estimate a typical viscous dissipation time
of stellar pulsations,
$t_{\rm diss} \sim \rho R^2 /\widetilde{\zeta}$, with $R \sim 10$ km
being the radius
of the neutron-star core.
The timescale $t_{\rm diss}$ is determined either by dissipation
in zone M ($t_{\rm diss}=t_{\rm M}$, if zones D and H are absent),
or dissipation in zone D ($t_{\rm diss}=t_{\rm D}$, if zone H is absent),
or dissipation in zone H ($t_{\rm diss}=t_{\rm H}$, if a representative
zone H is available). Then we obtain drastically
different dissipation times,
\begin{equation}
t_{\rm M}\sim {10 \, \omega_4^2 \over \widetilde{T}_9^6}~{\rm yr},
\quad
t_{\rm D}\sim {100 \, \omega_4^2 \over \widetilde{T}_9^4}~{\rm s},
\quad
t_{\rm H}\sim {10 \, \omega_4^2 \over \widetilde{T}_9^2}~{\rm ms}\;,
\label{dissipt}
\end{equation}
for the different cases.

For the subthermal pulsations, we have $\widetilde{T} = T$
and reproduce thus the estimates given by Haensel
et al.\ (\cite{hly02}).
For the suprathermal pulsations, $\delta \mu \gg T$,
the effective bulk viscosity and dissipation time 
(just as the relaxation times in Sect.\ \ref{supra}) 
stop to depend on temperature $T$ but depend on 
$\delta \mu$ instead, 
producing strongly enhanced nonlinear 
(in the pulsation amplitude, $\delta \mu$) 
viscous pulsation damping.
The damping will completely destroy the pulsation mode
if the dissipation time $t_{\rm diss}$ becomes comparable to
the pulsation period $2 \pi/\omega$. Thus, setting
$t_{\rm diss}=2\pi/\omega$ we can find the maximum pulsation amplitude
$\delta \mu_{\rm max}$ which a pulsating star can sustain.
For the pulsations in regimes M, D, and H, we obtain:
\begin{eqnarray}
    \delta \mu_{\rm max}^{\rm M}& \sim & 50 \, \omega_4^{1/2} \; {\rm MeV},
\label{deltaM} \\
    \delta \mu_{\rm max}^{\rm D}& \sim & 10 \, \omega_4^{3/4} \; {\rm MeV},
\label{deltaD} \\
    \delta \mu_{\rm max}^{\rm H}& \sim & 2 \, \omega_4^{3/2} \; {\rm MeV}.
\label{deltaH}
\end{eqnarray}
Accordingly, the viscous damping cannot allow the
star to pulsate with too large amplitude, and the maximum
amplitude does depend on the damping regime.
It is natural to assume that the strong damping
with the maximum amplitude
will rapidly reduce the pulsation amplitudes
to a lower level with a smaller damping rate.

\subsection{Superfluid matter}
\label{sect:small.sup1}

The effects of superfluidity are similar to those
described in Sects.\ \ref{sect:small.sup} and \ref{supra-sup}.
For subthermal pulsations, these effects
were considered, for instance, by
Haensel et al.\ (\cite{hly00,hly01,hly02}).
A strong superfluidity may drastically
(by many orders of magnitude) reduce
the bulk viscosity and the dissipation rate.
The effects of superfluidity on the bulk viscosity and 
the damping of 
pulsations 
in the suprathermal regime with $\delta \mu \ga \Delta$
are expected to be small, i.e., the results
of Sect.\ \ref{sect:normal.zeta.nonlin} can be used.

\section{Summary}

We have outlined
the most important factors which affect
the adiabatic index $\Gamma_1$ for density perturbations
and the viscous damping of pulsations in the neutron-star cores.
First, non-leptonic strangeness-changing processes
with hyperons may be rapid enough to establish
partial hyperonic equilibrium over pulsation periods
and decrease $\Gamma_1$ slightly below the
``frozen'' adiabatic index, $\Gamma_{\rm FR}$,
in hyperonic matter. In addition, they induce
a rapid viscous damping even in the subthermal
pulsation regime ($\delta \mu_j \la T$).
Second, suprathermal 
but still linear 
perturbations ($T \ll \delta\mu_j \ll \mu_j$)   may produce
a very rapid equilibration
in various relaxation channels which reduces $\Gamma_1$
to the fully equilibrium value $\Gamma_{\rm EQ}$,
enhances the bulk viscosity and damps the pulsations
quickly to lower amplitudes.
Third, superfluidity of baryons in the subthermal
regime increases relaxation times in various channels,
brings $\Gamma_1$ closer to $\Gamma_{\rm FR}$, and
reduces the viscous dissipation.

Therefore, a proper calculation of neutron star
pulsations and their dynamical evolution represents
a complicated problem. Much work is required to
study this problem in full detail. Our order-of-magnitude
estimates have to be replaced with accurate numerical
solutions of the equations of stellar pulsations
taking into account proper boundary conditions
and joint effect of various factors in all neutron-star
layers, from the surface to the center.
Our assumption of one typical pulsation
amplitude of chemical potentials, $\delta \mu$,
in Eq.\ (\ref{newT}) is an oversimplification.
In reality, one has to deal with the number
of {\it density dependent} 
amplitudes $\delta \mu_j$ for different
particle species $j$. The relaxation in different
equilibration channels and the viscous damping of pulsations
can be strongly nonuniform; one cannot exclude the
existence of thin layers in the neutron star cores,
where the damping is exceptionally strong.
For instance, they may be the layers where
new hyperons appear, with sufficiently small chemical
potentials $\mu_j$ just beyond their appearance threshold.
Neutron star pulsations in these layers may 
be suprathermal, producing enhanced damping.  
The problem of
suprathermal pulsation regime, without and with
superfluidity of baryons, is of special importance.
This regime will be accompanied
by huge energy release which will heat the star.

\begin{acknowledgements}
DGY is grateful to A.A.\ Pamyatnykh for a helpful discussion.
KPL and DGY acknowledge
hospitality of N.\ Copernicus Astronomical
Center in Warsaw.
This work was supported in part by the
RBRF (grants Nos. 02-02-17668 and 00-07-90183)
and KBN (grant 5 P03D 020 20).
\end{acknowledgements}

\end{document}